\documentclass{elsart}

\usepackage[latin1]{inputenc}

\usepackage{natbib}

\usepackage{amssymb}

\usepackage{url}

\journal{Ecological Economics}

\def\cal{\sf}

\date{7 March 2005}

\begin{document}

\begin{frontmatter}



\title{Thermodynamics and economics}


\author{Alastair D. Jenkins}
\ead{alastair.jenkins@bjerknes.uib.no}
\ead[url]{http://www.gfi.uib.no/\~{}jenkins/}
\address{Bjerknes Centre for Climate Research, Geophysical Institute,
Allégaten 55, N-5007 Bergen, Norway}

\begin{abstract}
The application of principles of thermodynamics and statistical
mechanics to
economic systems is considered in a broad historical perspective,
extending
from prehistoric times to the present day.  The hypothesis of maximum
entropy
production (MEP), which has been used to model complex physical systems
such
as fluid turbulence and the climate of the Earth and other planets, may
be
applied to human economic activity, subject to constraints such as the
availability of suitable technology, and the nature of political
control.
Applied to the current abundance of available energy from fossil fuel
reserves, MEP is shown to have significant policy implications.
\end{abstract}

\begin{keyword}
\def\sep{\unskip; }
Economic systems \sep
Statistical mechanics \sep
Thermodynamics \sep
Maximum entropy production principle \sep
Energy supply \sep
Political control

\end{keyword}

\end{frontmatter}

\section{Introduction}
\label{}
In this paper I discuss the applicability of concepts and techniques of
thermodynamics and statistical mechanics to areas within social science and
economics.  In particular, the concepts of entropy, and of entropy {\em
production}, are shown to be important.  Although the production of entropy,
in its information-theoretic sense of randomness, has been shown to be a
driving factor in the effectiveness of
markets (\citealp{MaasoumiE-RacineJ:je-2002-291}),
on the macro-economic scale of
whole societies it appears that a tendency for the maximisation of the
production of {\em classical thermodynamical entropy}, via the consumption of
available energy and other resources, may play a dominant role.  
The nature of economic
and social development and evolution will thus be determined by the physical
and other constraints which are placed on entropy production and energy
consumption.  The efficiency of systems of economic and political control will
depend on how well they take into account these underlying driving forces.
The application of the principles of thermodynamics and
information theory in economic systems is described 
by Ruth (\citeyear{%
RuthM:ee-1995-99,%
RuthM:jses-1996-125,%
RuthM:net-2005-243%
}), who states that `concepts and measures available from physics can be
used to improve our understanding of economic evolution if properly
placed into the context of socioeconomic processes'.  
In the present paper we consider the
thermodynamic and statistical mechanical aspects of economic processes, 
also from a historical perspective.

\section{Physical Background}
If we try to obtain complete solutions of the equations governing the 
mechanics of
a system with many degrees of freedom, we are faced, in general, with an
intractable problem.  A system  with as few as three degrees of freedom may be
subject to chaotic behaviour, in which a small perturbation in the initial
conditions will grow exponentially, so that the detailed long-term behaviour
is essentially unpredictable~(%
\citealp{LorenzE:jas-1963-130,%
KlavetterJJ:aj-1989-570%
}).
This intractability is an essential feature of macroscopic
systems which are composed of many atoms and molecules, one of the simplest 
type of which, {\em ideal gases}\footnote{%
Gases whose molecules effectively occupy only a negligibly small fraction of
the space they move about in.
}, being the subject of
intensive study during the 18th and 19th centuries.  The study of ideal gases
culminated in the kinetic theory of~\cite{MaxwellJC:cw-1859-377} and 
the statistical mechanics
of Boltzmann~(\citeyear{BoltzmannL:kawwmn-1872-275,%
BoltzmannL:kawwmn-1877-373%
}) and \cite{GibbsJW:epsm-1902}.

The basis of the kinetic theory of gases is that macroscopic properties such
as temperature and pressure are computable even though the motions of the
individual gas molecules may only be specified in statistical terms.  Indeed,
\cite{MaxwellJC:cw-1859-377} and
\cite{BoltzmannL:kawwmn-1872-275} 
showed that the components of the molecular velocity along any
coordinate direction are normally distributed with a variance proportional to
temperature and inversely proportional to the mass of the molecule, the
constant of proportionality $k$ being equal to what is now
termed {\em Boltzmann's constant}.

That gases have elegant mathematical properties was well known before the
development of kinetic theory.  In the mid-17th century, Robert Boyle
found that for a fixed mass of gas, the pressure and volume at a given
temperature were inversely proportional to each other.  Subsequent work by
Charles, Gay-Lussac, and Avogadro, led to the formulation of the ideal gas law
\[
pV = nRT,
\]
where $p$ is the pressure, $V$ the volume of a sample of $n$ moles\footnote{%
A mole is now defined as the quantity of a substance with the same number of 
molecules as the number of atoms in 12~grammes of carbon isotope 12.
}
of
gas, $T$ a suitable temperature scale (now the absolute temperature in degrees
Kelvin), and $R$ a universal constant.

Parallel to the development of gas theory there developed the theory of heat
engines.  On the basis that heat cannot flow spontaneously from a colder to a
warmer body, 
Sadi Carnot~(\citeyear{CarnotS-RPF-1824}) showed 
that a heat engine operating between two
temperatures could not be more efficient than a reversible heat engine
operating between the same temperatures, and that all reversible heat engines
operating between two given temperatures have the same efficiency.
This conclusion, together with Joule's demonstration of the equivalence of
mechanical work and heat, led, via the work of Kelvin and Clausius
(see \citealp{JaynesET-MEB-267}),
to the statement of the laws of thermodynamics:
\begin{description}
\item[{[First law:]}] Total energy is conserved: mechanical energy and heat are 
equivalent quantities;
\item[{[Second law:]}] There exists a state variable of a system (entropy).  The entropy
of an isolated system cannot decrease.
\item[{[Third law:]}] There exists a temperature (absolute zero, $T=0$) 
for which the entropy
of any system tends to a constant value (which may often be taken to be zero)
as it is approached.  
\end{description}
The rather mystical quantity {\em entropy\/} is, nevertheless, quite well 
defined.  For an ideal monatomic gas such as helium it is%
\footnote{This formula breaks
down at low temperatures: the Third Law of thermodynamics predicts $S\to0$
rather than $S\to-\infty$ as $T\to0$.  In reality, the gas would condense to
become liquid or solid at low temperatures.}
\begin{equation}
S = nR\left(\log {V\over n} + {3\over2} \log T + \hbox{constant}\right).
\label{gas-entropy}
\end{equation}

Useful thermodynamic relations to be aware of are:
\[
\Delta E = Q + W,
\]
which represents the First Law of thermodynamics, $\Delta E$ being the change
in total energy of the system due to a supply of heat $Q$ and mechanical work
$W$; and
\begin{equation}
dE = T dS - P dV, \label{ftr}
\end{equation}
which is the {\em fundamental thermodynamic relation}, valid for both
reversible and irreversible changes of state, $dE$, $dS$, and $dV$ representing
infinitesimal changes in energy, entropy, and volume, respectively.  To obtain
finite changes, the relation~(\ref{ftr}) should be integrated between the
initial and final states: 
\[
\int_{\rm initial}^{\rm final}dE = \int_{\rm initial}^{\rm final}T dS - \int_{\rm
initial}^{\rm final}P dV.
\]

We may also employ the following formula, due to Clausius:
\begin{equation}
dS \geq dQ/T, \label{eq-clausius}
\end{equation}
with equality where the change in the state of the system is reversible.
This relates the change in entropy of a system to the amount of heat supplied
to it.  We see that if the heat flows into or out of a system at a high
temperature, it produces
a smaller change in the entropy of the system than if it flows in or out at a
low temperature. 

It may also be useful to employ {\em thermodynamic potentials}, such as the {\em
enthalpy}
\[
H = E + PV, \qquad dH = T dS + V dP,
\]
when considering changes at constant pressure, or the {\em Gibbs free energy}
\begin{equation}
G = E - TS + PV,\qquad dG = -S dT + V dP,
\label{GFE}
\end{equation}
for changes at constant pressure and temperature (e.g.~phase changes such as
melting and evaporation).

\subsection{Non-thermal entropy changes}
The entropy of a system may also change (increase) in processes where there
are no thermal effects, for example, during the mixing of two different gases
which do not react with one another.  Consider two adjacent compartments $C_1$
and $C_2$, separated by a partition, each
of volume $V/2$, containing $n/2$ moles of 
monatomic gas species 1 and 2, respectively.
By~(\ref{gas-entropy}), the entropy of the gas in each compartment is $S_i =
(n/2)R[\log (V/n) + (3/2) \log T + c_i]$, where $i$ is either 1 or~2, the $c_i$
being the relevant constant in~(\ref{gas-entropy})~for each gas.  The total
entropy is thus
\[
S_{\rm initial} = S_1 + S_2 = nR\left[\log {V\over n} + {3\over2} \log T
+ {1\over2}(c_1+c_2)\right].
\]
If we then open the partition between the compartments and let the gases mix,
keeping them at the same temperature, each gas species will then occupy volume
$V$, and the total entropy will become
\[
S_{\rm final} = S_1 + S_2 = nR\left[\log {2V\over n} + {3\over2} \log T
+ {1\over2}(c_1+c_2)\right] 
= S_{\rm initial} + nR\log 2.
\]
This is a elementary example to show that entropy is a measure for the 
disorder of a system.  

\subsection{Relation between thermodynamics and microscopic dynamics}
The relation between
entropy and the microscopic dynamics of an ideal gas was clarified by Ludwig
Boltzmann~(\citeyear{BoltzmannL:kawwmn-1872-275}), 
who showed that the following quantity would always
decrease in a thermally and mechanically isolated system:
\[
{
\cal H} = N \int\!\! f({\bf c})\log f({\bf c})\,d{\bf c},
\]
where the vector ${\bf c}$ is the gas-molecule velocity, and $f({\bf c})$ is
its probability density.  The entropy per unit volume turns out
to be $-k{\cal H}$, up to an additive constant.  Subsequently,
Boltzmann~(\citeyear{BoltzmannL:kawwmn-1877-373}) generalised this result to his
famous formula
\begin{equation}
S = k\log W,
\end{equation}
where $W$ is a measure of the volume occupied by the dynamical system
comprising the ideal gas molecules, in the phase space of $6N$ dimensions,
whose coordinates are the components of position and momentum of all $N$
molecules. 

The fact that ${\cal H}$ always decreases appears to be paradoxical, as the
equations describing the dynamics of the gas molecules and their
interactions (elastic collisions) are time-reversible (Loschmidt's
Paradox).  In Boltzmann's
lifetime, this was regarded as a serious defect in his theory.  Boltzmann
himself made the assumption of molecular chaos ({\em Sto{\ss}zahlansatz}), that
is, that before the collision of any two gas molecules, the molecules'
velocities are uncorrelated, and it is from this assumption that we may derive
a time-asymmetric equation for the evolution of the molecular velocity
distribution (the Boltzmann equation) from the time-symmetric equations of
dynamics.  The validity of Boltzmann's approach for practical purposes has been
confirmed by the fact that it is able to predict, very accurately, using
Enskog's~(\citeyear{EnskogD:KTG-1917}) perturbative solution and related 
approaches
(\citealp{%
ChapmanS_ptrs-1916-279,%
ChapmanS-CowlingTG:MTNUG-1970%
}), 
many of the transport properties of gases, such as
viscosity, thermal conductivity, diffusion coefficients, and also to predict
quantitatively more non-intuitive effects such as thermal diffusivity
(the tendency for a concentration gradient to be set up in a gas mixture when
there is a temperature gradient).  More recently, it has been shown that
such apparently time-asymmetric behaviour can arise from time-symmetric
equations used in simulations of molecular dynamics, as a result of the
presence of both repelling and attracting subsets of the phase space
(\citealp{%
NoseS-jchp-1984-511,%
HolianBL-HooverWG-PoschHA:prl-1987-10%
}).
 
\subsection{Mixtures of substances}
The thermodynamics of mixtures of substances was put on a firm mathematical
footing by J.~Willard Gibbs~(\citeyear{GibbsJW:EHS-1875}), 
who introduced the concept of 
{\em chemical potential\/} $\mu$.  For a system with $N$ different substances, 
equation~\ref{ftr} is replaced by
\begin{equation}
dE = T dS - P dV + \sum_{i=1}^N \mu_i dn_i, \label{ftrchem}
\end{equation}
where $n_i$ is the number of moles of substance $i$, which has a chemical
potential $\mu_i = (\partial G/\partial n_i)_{T,P}$, the partial derivative of
the Gibbs free energy (see Eq.~\ref{GFE}).  
For substances participating in
chemical reactions, chemical equilibrium is reached when the sum of chemical 
potentials of the reactants is the same as the sum of the chemical potentials
of the products.

\subsubsection{Statistical mechanics of mixtures}
Gibbs also laid the foundation for the statistical
mechanics of general systems in thermodynamic
equilibrium (\citealp{GibbsJW:epsm-1902}).  His formulation was sufficiently
general that it remained valid with the development a quarter-century later of
quantum mechanics.  The general principle behind Gibbs' theory is
essentially the maximum-entropy argument devised by Boltzmann,
but applied to dynamical systems more general than those
describing the behaviour of ideal gases, and
with additional consideration of the presence of molecules of different
chemical composition.  From maximising the number of ways the phase space
of the whole system may be occupied, subject to the system having its
given energy, momentum, and composition, Gibbs showed, in the limit of
large particle number, that the behaviour of a system in thermodynamic
equilibrium could be described in terms of what is now called the
{\em grand partition function\/}
\[
Z = \left\langle\exp\left[
\left( -E +\sum_{j=1}^m \mu_i n_i
\right)\biggm/(RT)
\right]\right\rangle,
\]
where the angle brackets represent the average or mathematical expectation.
Gibbs' theory, although at the time expressed in terms of the laws of classical
mechanics, is also applicable to systems which obey the laws of quantum
mechanics, developed in the 1920s by Heisenberg, Dirac, and Schrödinger, among
others, where the positions and momenta of the components of the system must be
expressed as abstract linear operators rather than as numbers.

\subsection{Non-Equilibrium Systems}
So far, we have only considered quantitatively the behaviour of systems in
thermodynamic equilibrium.  A mathematical theory of a non-equilibrium
phenomenon---that of heat conduction---was developed by \cite{FourierJBJ:mars-1826-153}.
In microscopic terms, for a gas, the heat is carried and 
transmitted via the random molecular motions, the quantitative theory being
developed by Maxwell and Boltzmann for a specific case of the
intermolecular potential (proportional to the $-5$~power of the distance
between the molecules), and by Chapman and Enskog for general
intermolecular force laws.  Heat conduction in solid substances is via a
similar mechanism, the role of the gas molecules being played by {\em
phonons}, quantum-mechanical particles which are associated with elastic
vibrations of the material.  

A general quantitative theory of non-equilibrium
thermodynamics was developed by \cite{%
OnsagerL:pr-1931-405,%
OnsagerL:pr-1931-2265%
}.  Onsager's theory is
valid for systems which are close to thermodynamic equilibrium, in the sense
that there is a linear relation between {\em thermodynamic forces\/} 
(such as the temperature gradient) and {\em flows}, such as the heat flux.  In
addition to heat conduction and molecular diffusion (in solids and liquids as
well as gases), Onsager's theory covers phenomena such as the thermoelectric
effect (heating or cooling caused by the passage of an electric current
between two different materials), thermal diffusion (the diffusive separation
of different substances in the presence of a temperature gradient), and so on.

For systems far from thermodynamic equilibrium, the general statistical
mechanical theory has until recently been incomplete.  It has been observed that there is often a
general tendency for the rate of entropy production to be maximised.  For
systems in a (stochastically) steady state which maintain a near-constant
temperature, this is equivalent to saying that the rate of energy dissipation
is maximised.  

The hypothesis of maximum entropy production (MEP) has been applied with some
success within the field of fluid mechanics, providing some quantitative
results relating to the properties of the notoriously intractable problem of 
turbulent fluid flow.  Busse (\citeyear{Busse:jfm-1970-xxx}) and Malkus
(\citeyear{MalkusWVR:jfm-1956-v1}) found some solutions to the
hydrodynamic equations for a shear flow which maximised the viscous energy
dissipation.   Although these solutions, involving complex stationary flow
patterns, were not strictly turbulent, they do give rates of cross-flow
momentum flux (equivalent to turbulent shear stress) which are remarkably
close to those observed in laboratory experiments and in fine-scale
time-dependent numerical model simulations.  A similar approach has also been
applied with some success in the study of thermal convection%
.

On a more ambitious level, the MEP hypothesis has been applied to the even
more complex system of the global climate
(\citealp{%
PaltridgeGW:qjrms-1975-475,%
PaltridgeGW:nature-1979-630,%
OzawaH-OhmuraA:jc-1997-441,%
PujolT-FortJ:tellus-2002-363%
}).  This work
has been applied to the  ocean thermohaline circulation
climate system (\citealp{ShimokawaS-OzawaH:qjrms-2002-2115}),
and Lorenz {\em et al.} (\citeyear{LorenzR-etal:grl-2001-415}) have shown
how MEP
may be applied to make quantitative predictions of the climate of other
planetary bodies (Mars, Venus, Titan).  An overview of the application of
MEP to turbulence and climate-related studies is given by \cite{%
OzawaH-ShimokawaS-SakumaH:pre-2001-026303%
}.

Theoretical progress in understanding the concept of maximum entropy in
non-equilibrium systems was made by 
Dewar~(\citeyear{DewarRL:jphysa-2003-631}), 
who showed that MEP in a
system in a steady state would be attained by the maximisation of {\em path
information entropy} 
\begin{equation}
S_I = -\sum_\Gamma p_\Gamma \log p_\Gamma,
\label{eq-pie}
\end{equation}
where the summation index (or integration variable) $\Gamma$ is over
possible paths in the phase space of the dynamical system, and the
$p_\Gamma$ are the probabilities of the system will follow the individual
paths.  Equation~\ref{eq-pie} was first employed by
\cite{JaynesET:pr-1957-620}, employing the information-theory entropy
concept of \cite{ShannonCE:bstj-1948-379}. In the case of systems in
thermodynamic equilibrium, it reduces to the statistical-mechanical
definitions of entropy deduced by \cite{BoltzmannL:kawwmn-1877-373} and
\cite{GibbsJW:epsm-1902}, but may also be applied to systems not in
equilibrium.  It should be noted that the paths $\Gamma$
in~(\ref{eq-pie}) are restricted to those which are dynamically
realisable: for example, a system constrained initially to be in
a non-equilibrium state will tend to equilibrium with a finite relaxation
time, given, for example, by the dynamics of intermolecular collisions.
This principle of maximal $S_I$ has been shown to reproduce the results
of Onsager's theory, and is also capable of making predictions of the
behaviour of systems far from equilibrium (\citealp{%
JaynesET:mef-1979-15,%
RobertsonB:pr-1966-151,%
RobertsonB:pr-1967-175,%
RobertsonB:PP-1993-251%
}).
A review of the application of the theory of dynamical systems to
non-equilibrium statistical mechanics is given by
\cite{RuelleD:arxiv-chao-dyn-9812032}.

The above principle of maximum of path entropy may be used to explain, in
addition to processes involving heat flow, such phenomena as the behaviour of
systems subject to critical behaviour, for example, a sand
pile (\citealp{BakP-TangC-WiesenfeldK:prl-1987-381}).  The MEP
property of networks exhibiting self-organised criticality has been applied by
\cite{LorenzR:gra-2003-12837} 
to economic market systems, where the profit (difference between
buying and selling price) realised by a participant in the market plays the
part of energy dissipation or entropy production.  Lorenz's idea can be
thought of as a microeconomic application of MEP.  On the macro-economic level
of whole economies, I contend that MEP may be applied using the {\em usual
thermodynamic definition\/} of entropy.  To understand how this is so, I will
outline human economic activity within its historical context.

\section{Human Activity}

The distinctive contribution of human activity to the global thermodynamic
balance comes with the exploitation of fire.  Before that time, human
influence was effectively indistinguishable from that of other forms of living
organism.  Living systems tend to have a high degree of order (negative
entropy), but they maintain this state by `feeding' on energy and mass 
sources with a
low specific entropy, and producing waste products with a higher
entropy.  For example, plants utilise radiant energy from the Sun, with
an effective temperature of over 5000\,K, and release the energy as heat at an
ambient temperature of about 300\,K.  From Eqs~\ref{ftr}
and~\ref{eq-clausius}, we see that the plants will tend to produce entropy. 
Similarly, the entropy of food which animals consume is less than the entropy
of the waste products which they produce.  It also appears that the
appearance of vegetation on the planetary surface tends to reduce both
the temperature and the albedo of the surface, thus increasing the
entropy production rate for a given input of solar radiation (\citealp{%
UlanowiczRE-HannonBM:ProcRoySocB-1987-181,%
SchneiderED-KayJJ:mcm-1994-25,%
KleidonA-FraedrichK-HeimannM:cc-2000-471,%
KleidonA-LorenzR:net-2004-xxx}).

Although naturally-occurring fires, caused, for example, by lightning strikes,
have always existed, the generation and exploitation of fire by human activity
has increased entropy production by enabling a more rapid dissipation of the
free energy available from organic carbon and atmospheric oxygen, and the
distribution of the effects of fire over a wide area, for example, to increase
the availability of game for hunting (\citealp{BeatonJM:aroc-1982-51}).
Indeed, the supply of firewood has been a significant limitation on human
economic activity, in historical periods, for example, in England in the
fifteenth and sixteenth centuries (\citealp{LeeJS:ehr-2003-243}),
and in the present day in semi-arid mountain 
regions where the main economic activity is subsistence farming and
grazing (\citealp{EckholmE:wp-1975-2}).

Entropy is a concept which, in addition to being a thermal property of
materials, is also a property of the distribution of the materials.  The
entropy of a fixed mass of gas, for example, depends on the volume it occupies
(see Eq.~\ref{gas-entropy}).
The same applies for substances in
solution (\citealp{DebyeP-HueckelE:pz-1923-305}) or in mixtures.  
A prehistoric example of entropy production
by this mechanism is given by the exploitation of flint resources---the
production of tools, such as in the flint-mining areas of Norfolk,
U.K. (\citealp{BarberM-FieldD-ToppingP:nfme-1999}).
A concentrated supply of
flint is broken up and dispersed, as useful objects which are eventually
discarded, and also as waste material.

The development of agricultural techniques enabled the increase by digging of
the entropy of the soil and the release of mineral resources for plant growth,
whose subsequent depletion coupled with an increased human population led to a
rapid extension of the agricultural frontier, a process which took place in
Europe, Asia, and Africa in prehistoric times, and which was repeated in the
Americas and Australasia in more recent historical periods.  The expansion of
agriculture is an early example of available free energy contributing both to
direct entropy production and by `investment' in `entropy-productive capacity'
in the form of forest clearance and other activities which enable future
increase in entropy production.

Agricultural techniques enable a greater production of `fuel', for humans,
livestock, and `pests', than would otherwise be possible.  The necessary
increased absorption of sunlight for photosynthesis will decrease the Earth's
albedo correspondingly, increasing the energy absorbed (and thus the entropy
produced) from solar radiation (\citealp{KleidonA-FraedrichK-HeimannM:cc-2000-471}).  This will be the case both for when the
original landscape is forested, and when the original landscape is arid and
subsequently irrigated for food production.

The production of tools by smelting of metal ores (copper, tin, iron),
although it produces end products of lower entropy than the original raw
materials, will necessarily produce entropy in the waste products.  The tools
produced will increase entropy production via the production of food, and
their use in human conflicts will increase entropy in a more `disorderly'
manner.  Eventually the tools will corrode, into substances of a similar
specific entropy to the original ores, but their distribution will be less
concentrated and thus have greater entropy.

Societies based on agriculture may of course reach considerable complexity and
sophistication.  The `entropy production' of an agricultural society at a
state in which the population is more-or-less constant
(\citealp{LaslettP:wwl-1971}) may be
maximised by the export of food and excess population to urban areas (which
have higher death rates).  Social stratification may enable a more `effective'
entropy production.  Laslett also states (\citeyear{LaslettP:wwl-1971}, p.~66) that
the controlling {\em gentry\/} in 17th century England `pressed, like the
atmosphere, evenly, over the whole face of England', that is, as in a
body of gas evenly distributed throughout a container, their spatial
distribution was in a state of maximal 
entropy.  Conflicts
with other societies provide an additional source of energy dissipation~/
entropy production and absorption of `excess population'.

\section{Industrial Societies}

Of course, the so far greatest expansion of anthropogenic entropy production,
to an extent which will almost undoubtedly affect the global climatic balance,
has occurred after the start of the Industrial Revolution.  Large-scale iron
production was previously limited by the timber available for charcoal-making,
for example in the Caledonian
Forest of Scotland (\citealp{%
TittensorRM:sf-1970-100,%
DyeJ-KirbyM-etal:soha-2001%
}).
The scene was then set for the conversion of the huge geological reserves of
coal and, later, petroleum, laid down over the past hundreds of millions of
years, to atmospheric carbon dioxide (CO$_2$), at an unprecedented speed.  The
fact that this process could not happen instantaneously is due to the energy
investment necessary for coal mining, ore extraction, smelting works,
transport infrastructure (such as railways) and other facilities and processes
necessary for society to absorb the energy and materials generated.  The large
investment of energy required to construct and develop mines and processing
resources limits the rate of increase of total production and consumption of
coal. It can probably be shown that the coal production during the
Industrial Revolution, in Britain, for example, increased at a rate which
maximised the total entropy production, subject to the technological
constraints which applied at the time.  For example, we may assume that if the rate of coal
extraction (in units of the energy available by burning the coal) is $P$, the energy used
in coal extraction and distribution is $\alpha P$, and the power used
to increase production capacity is $hP$.  Furthermore, we assume the following relation
between the rate of change of coal production and the investment in increasing production
capacity:
\begin{equation}
{dP\over dt}=\beta h(2h_0-h)P - hP.
\end{equation}
The rate of entropy production associated with the burning of $Q$ units of coal is $Q/T_a$,
where $T_a$ is the the ambient temperature.  Under the assumption that $T_a$ is
constant, the entropy production will be maximised for $h = h_0 - 1/(2\beta)$, and the
rate of coal extraction will increase exponentially\footnote{It may be argued that a higher
rate of entropy production would be attained just by burning coal `in place' instead of
distributing it for `productive' use.  However, not only is the human socioeconomic system
not designed for such an `unprofitable' activity (although a certain amount of spontaneous
combustion does, in practice, unavoidably take place), but also if the coal is distributed
to the wider economy it will enhance the consumption of other raw materials, and thus
produce extra entropy at a rate proportional to the product of their chemical potential and
their consumption rate (see Eq.~\ref{ftrchem}).}.

The development of efficient transport systems such as railways tended to
reduce the entropy production during the transport process itself.  However, by
increasing the total transport capacity and making the wider distribution of 
products possible, the total entropy production rate of the whole society and
economic system would be increased, that is, economic growth would be
stimulated.  The effectiveness of the railways in the economic development of
the U.S.A. in the 19th century was nevertheless disputed 
by~\cite{FogelRW:RAEG-1964}, who claimed that a combination of water and road
transport would have been almost as effective.  However, his conclusion was
disputed by~\cite{HolmesTJ-SchmitzJA:frmqr-2001-3}, who pointed out that if the
waterways had not faced rail competition, they would have been subject to a
greater extent of restrictive practices, such as a tendency for individual groups, such
as dockers' unions, to
maximise their own entropy production (see also \citealp{RuttenA:IR-2003-285}).

Passing on to present-day society, oil and gas are supplanting coal as they
require the dissipation of less energy to produce and transport---thus their
actual production (and the consequent entropy produced as they are consumed)
may be greater than if coal was still used as fuel.  Coal production thus
tends to be concentrated in the reserves which are least energy-intensive to
extract and transport to the markets, the other less `economic' mines facing
closure.

Today's abundant sources of fossil-fuel energy result in a situation whereby
the tendency to maximum entropy production becomes highly visible.  Transport
provides a good example.  Although for heavy bulk cargo the greater energy
efficiency of marine and rail transport still gives an advantage to these
modes, the tendency for MEP leads to the dominance of the modes with higher
energy dissipation: road, and, increasingly, air transport.  That rail has, in
some countries, not entirely disappeared as a mode of medium-to-long-distance
passenger transport, lies perhaps in the fact that the generally greater
comfort of rail as opposed to bus transport leads to rail being comparably or
even less energy-efficient per passenger-kilometre 
(\citealp{%
AndersenO-UusitaloO-etal:etg-1999,%
AndersenO:VF-2001-5}).

In metropolitan areas, there is a tendency towards the use of rail for
passenger transportation, as it has a greater carrying capacity than road
transport, and thus allows for higher levels of economic activity (and entropy
production).  Cars standing in traffic jams, although they may emit
considerable amounts of noxious substances, consume relatively little fuel and
thus make a smaller contribution to entropy production.

It may be thought that the command economies of the former Soviet bloc countries
provide a counterexample to the tendency for MEP.  These economies, with their
emphasis on coal/steel-based heavy industry and the consequently large energy
consumption and entropy production, are being superseded by free-market
economies which have somewhat lower energy consumption.  However, we can
consider this phenomenon as a result of the relaxation of a constraint---the
entropy production in the part of the society outside the industrial complex
was less than it would have been in the absence of the rigid system of
political control.  Indeed, the economies of the former Eastern bloc countries
are now showing MEP characteristics of free-market economies, such as reliance
on private motor vehicles, rather than less energy-intensive public
transportation (\citealp{HanellT-BengsC-etal:nr-2000-10}).

Modern agriculture has increased its entropy production above what could be
achieved in the pre-industrial age.  The availability of cheap extractive and
transportation technology has enabled the transport of minerals such as
phosphate to agricultural areas.  The fixation of nitrogen from the air to make
it available for fertilising plant growth may be performed industrially, via
chemical engineering technology such as the Haber process for producing
ammonia (\citealp{SmilV:ee-2004}), thus short-circuiting the slower naturally-occurring bacterial
processes.

The heavy investment required for industrial development has given rise to a
tendency for `business cycles', a phenomenon recognised from Victorian
times (\citealp{DuesenberryJS:bs-1958}). 
The price depression caused by increased capacity and competition then leads to
reduced profits for investors.  Political forces may then arise, with the aim
of maintaining employment, leading to a tendency for the provision of
subsidies, or the imposition of tariff barriers to restrict competition.  This
political process can be thought of as a means to maintain the entropy
production of the industrial process concerned, although the wealth (money)
production in the economy as a whole will not be optimal.

The recent `dotcom bubble' may also be interpreted in terms of entropy
production.  The amount of bits of information produced in the information technology industry
is vast, but its `Shannon entropy' is negligible compared with
the associated entropy production in industrial production, 
transportation, building, and so on.  Even though many investors lost a lot of
money, the industry they invested in generated considerable entropy.

There has been recent discussion of the benefits of
free and open source computer software, which may be most efficient in the 
long  run, since there is a greater potential to build new and more useful
software products in the absence of restrictions on sharing computer source
code (\citealp{MustonenM:cl-2003}).
At present, however, 
there is much more money (and thus entropy) produced in the market for
traditionally-produced commercial 
software products.  The same condition applies to other forms of 
intellectual property, such as books, music, and so on.
The owners of copyrights and patents produce entropy (for themselves) 
`while the sun shines'.

\section{Environmental Implications}

Tendency for maximisation of entropy production leads to rapid use of available
resources, and a general tendency for increased environmental pollution.  
The rapid use of fossil fuel reserves, built up over hundreds of millions 
of years,
may very well produce a serious greenhouse-gas-induced climate change before 
their eventual exhaustion.  It will be a challenge for the world's political
system to impose suitable constraints, particularly since the potential for
entropy production from renewable resources such as solar and wind energy is at
first sight quite restricted.

What appear to be far-fetched ideas for storing waste CO$_2$ in subterranean
storage reservoirs, such as depleted oil and gas fields, may in fact be
realistic from an entropy-production viewpoint.  A substantial proportion of
the energy produced will need to be consumed in compression and transport of
CO$_2$ to the reservoirs, but this, of course, will lead to an increase in
entropy production, which may be favoured by economic and political
forces.

\section{Implications for Economic Planning}

We have seen in the previous sections how human activity, as well as the Earth's
geophysical and biological processes, may act in a way to produce entropy at a maximum rate,
subject to practical constraints such as the availability of resources and other raw
materials, and the energy which must be diverted in order to extract them.  It is
consistent with the idea that economic growth will be maximised if the economic system is
subjected to as little disturbance as practicable: rigid economic planning is liable to
reduce the rate of growth, even if the plans do not intentionally have this effect.
Another implication of the MEP hypothesis is to suggest an alternative, perhaps more
objective, alternative to the concept of economic {\em utility}.  Instead of individuals
acting to maximise their own, highly subjective `utility', which is expressed in the market
in monetary terms, we may hypothesise that individual and political pressures expressed in
society as a whole tend to maximise entropy production.  This again suggests further how
political power in a society is not merely dependent on the financial resources available
to individuals or corporations, but is alternatively a function of the amount of energy
and material 
resources they control for entropy production.  We would therefore expect that corporations
which control access to energy reserves, and also governments which regulate the allocation
of such reserves, should exert more political and economic power, even if in financial
difficulty, than those without such access and control.  The deliberate addition to motor
gasoline of the toxic chemical tetraethyl lead, and the consequent astronomical
increase of lead deposition throughout the global environment, is a
case in point (\citealp{MilbergRP-LagerwerffJV:jeq-1980-6}). 

It may thus be seen that the tendency for MEP due to human economic activity may not in all
respects have beneficial effects.  However, this MEP tendency is subject to the constraints
which are either originally present or which may be imposed on the system.  The 
economic systems in different parts of the world may act as a guide.  The U.S.A. has the
world's greatest entropy production {\em per capita}, and this is associated with a
relatively unconstrained economic system.  However, even in the U.S.A., uncontrolled
economic activity is regulated by such means as local building codes, and national social
security and environmental protection schemes.  In Europe, the rather lower per capita
entropy production is perhaps associated with a higher level of central infrastructure
planning and social welfare levels, and manifests itself by means of a rather smaller
spread of individual income, a somewhat greater use of public transportation,
{\em et cetera}.

However, economic regulation may be driven by political pressures which act to increase
entropy production.
The high social welfare and public infrastructure spending in the Nordic countries
(\citealp{DowrickS:ej-1996-1772})
 may be a
response to those countries' generally low population density: entropy production may be
maximised by keeping the population spread throughout the country rather than yielding to
a natural centralising tendency.  The agricultural subsidies applied in the European Union
(\citealp{deGorterH-MeilkeKD:ajae-1989-592})
and elsewhere will also have a similar effect.

The Montreal protocol (\citealp{MurdochJC-SandlerT:jpe-1997-331}) 
on the production of chlorofluorocarbons (CFC) and other 
substances that deplete the stratospheric ozone layer can also be seen to be 
a MEP-driven agreement.  When it was eventually realised that stratospheric ozone levels
were being reduced significantly by CFC-catalysed chemical reactions in the atmosphere, and
that the resulting levels of solar ultraviolet radiation were likely to increase, leading
to an increased danger of skin cancer and possible adverse effects on oceanic plankton
populations, the consequences to the tourist and fishery industries became clear.  The
possible reduction in economic activity (entropy production) in these industries helped to
generate the political pressure which led to a successful agreement.  The proposed
reduction in CO$_2$ releases envisaged in the Kyoto Protocol
(\citealp{BabikerMH-JacobyHD-etal:esp-2002-195})
 can then be seen to be much
more difficult to attain, as they will lead to a {\em reduction\/} in global entropy
production.  If air transport is included in the CO$_2$ emission reduction requirements,
this will obviously be vigorously resisted by the tourist industry.  In the medium term, an
entropy production increase may be achieved by sequestration of CO$_2$ emissions, in
geological formations, the ocean (\citealp{%
DrangeH-HauganPM:ecm-1992-697,%
BrewerPG-FriederichG-etal:sci-1999-943%
}), or elsewhere, a process which will require a significant
fraction of the global total energy production.  This, however, will require a large
investment in processing capacity, in
energy as well as in financial terms. 

Model studies of the changes in industrial processes induced by
different types of carbon emission reduction incentive have been
performed by \cite{RuthM-AmatoA-DavidsdottirB:es-2000-269} and
\cite{RuthM-AmatoA:ep-2002-541} for U.S. iron and steel production, and
by \cite{RuthM-AmatoA-DavidsdottirB:es-2002-119} for U.S. ethylene
production.  In these studies, it was found that the response of the
industries to different types of incentive varied: a CO$_2$ tax was not
so effective in reducing emissions as policies, such as research and
development stimuli, which were specifically directed at the industries
concerned.  The latter type of policy may be regarded as `catalytic' in
that it affects the `reaction kinetics' or dynamics rather than
affecting the state of economic equilibrium to be attained.  In general,
to influence industries and economic processes to direct them towards
specific environmental goals, it is necessary to implement policies
which both encourage a transition towards the specified goal, which must
also be `thermodynamically' realistic, and also ease the constraints
which may inhibit such a transition.

It can thus be seen that the long-term future of human society, as affected by greenhouse
gas induced climate change, provides a serious political challenge to the process of
economic and social planning.  The consequences of any imposed 
technical or economic regulations which purport to protect the global or local climate and 
environment should be assessed realistically, not only with regard to their ecological
consequences and financial
implications for the stake-holders, but also with regard to their implications for energy
production and cycling and for entropy production.  The entropy production factor may be
used as a proxy for as-yet-unforeseen economic and political forces.

\section{Concluding Remarks}

In this essay we have introduced concepts from thermodynamics and statistical mechanics,
for both equilibrium and non-equilibrium systems, which may have applicability for the
global economy and society generally.  The system which contains the Earth's climate,
biosphere, human society and economy, is not at equilibrium, but is an open system,
characterised by an energy flux from solar radiation which is re-radiated to outer space.
Human activity, in addition, produces more energy 
from fossil
fuels, 
which is also re-radiated, 
and also extracts minerals from ores and re-distributes their products.

Thermodynamic systems which are isolated have a tendency to reach equilibrium, a state of
maximum entropy, subject to constraints such as the total energy and the chemical
composition.  For open systems, at least those which are on average in a steady state, 
there has been observed a tendency for maximisation of entropy {\em production\/} (MEP), a
hypothesis which has been proved mathematically, under certain circumstances, by
\cite{DewarRL:jphysa-2003-631}. This MEP tendency has been observed in the dynamics of the
climate of the Earth and other planets, in fluid turbulence and convection, in the
dynamics of network systems subject to self-organised criticality,
and in the
climatic effects of the biosphere, which suggests a `rational' explanation of the `Gaia'
hypothesis that the Earth's climate--biosphere system is self-regulating
(\citealp{LovelockJ-MargulisL:Tellus-1974-2}).

It should be noted that alternative, related hypotheses for the
behaviour of complex mechanical, thermodynamic, biological and economic
systems have been put forward.  One example is that of maximum
utilisation of {\em available energy\/} or {\em exergy} (\citealp{%
EdgertonRH:AEEE-1982,%
KayJJ:SOLS-1984,%
SchneiderED-KayJJ:mcm-1994-25,%
NielsenSN-UlanowiczRE:em-2000-23,%
FraserR-KayJJ:TRSLSP-2002-xxx%
}).  The quantity `exergy' may be defined
as the maximum energy available to a system by the operation of an ideal
heat engine which is able to exchange heat at the ambient environmental
temperature.  Exergy is said to have simpler conservation laws than
(non-conserved) entropy, and to be easier to compute.  Unlike entropy,
the computation of exergy relies on the presence of an ambient heat
bath.  The concept of maximising entropy production should thus be of
more general applicability than that of maximising exergy utilisation,
although the two concepts will be equivalent under suitable
circumstances.

In this paper I assume that MEP applies to human economic activity, the `entropy' being
the usual thermodynamic entropy.  This has the consequence that macro-economic processes are
directly related to available resources of energy, and also of minerals and other raw
materials.  In today's society, with abundant energy available from fossil fuel reserves,
economic activities, for example, modes of transport, are favoured if they consume more
energy, and thus produce more entropy, than their alternatives.  International agreements,
even those which purport to restrict economic activities, such as the 
Montreal 
protocol (\citealp{MurdochJC-SandlerT:jpe-1997-331}), will nevertheless be favoured politically if they will lead to increased entropy
production, or, at least, will avoid the risk of reduced entropy production.

The major challenge facing world society today is that of global climate change induced by
the emission of greenhouse gases such as CO$_2$.  Efforts, such as the Kyoto protocol, to
limit or reduce greenhouse gas emissions, face severe political difficulties, since they
will result in limits to entropy production.  In order for such agreements to succeed, it
must become clear, to each individual as well as to the political, intellectual, and
commercial élite, that severe economic and social consequences will ensue, to the extent
that even the foreseen `entropy production' (in the way of, for example, social
interaction, tourism, sport, and entertainment) will {\em decrease}, if an agreement to
limit emissions is not made.  In order for everyone to come to such a realisation, a
substantial investment must be made, in the monitoring of climatic variables in order to
detect early signs of adverse climate change, in research, both for understanding the
present and past climate and climate--biosphere interactions, and for modelling and
predicting future climate change and its effects.  Finally, it will be necessary to 
put into place a rational, believable,
and well-thought-out campaign of publicity and education, so that there is a general
understanding of the prospects for climate change and its consequences.  This, one may
think, would be an overwhelming task, but the success of such (substantial entropy
producing!) organisations such as {\em Greenpeace\/}\footnote{{\em Greenpeace\/} gives
information on its (entropy-producing) ships at URL \url{http://archive.greenpeace.org/ships.shtml}.} 
shows that it is possible to mount a
successful challenge to entrenched economic interests.

\section*{Acknowledgements}
This work was performed with support from the Research Council of Norway.
I thank Joachim W.~Dippner and Martin Miles for useful and stimulating
discussions.
This is paper no.~0000 of the Bjerknes Centre for Climate Research.  


\bibliographystyle{elsart-harv}
\bibliography{alastair_abbrevs,alastair}





\end{document}